\documentclass[twocolumn]{revtex4}
\usepackage{natbib}
\usepackage{graphicx}

\begin{document}

\title{Application of van der Waals Density Functional to an Extended System: 
Adsorption of Benzene and Naphthalene on Graphite}

\author{Svetla D. {Chakarova K{\"a}ck}, Elsebeth Schr\"oder, Bengt I. Lundqvist\\}
\affiliation{Department of Applied Physics, Chalmers University of Technology,
SE-412 96 {G\"oteborg}, Sweden}
\author{David C. Langreth\\}
\affiliation{Department of Physics and Astronomy, Rutgers University, 
Piscataway, N. J. 08854-8019, USA}

\date{\today}

\begin{abstract}  
It is shown that it is now possible to include van der Waals interactions
via a nonempirical implementation of density functional theory to describe
the correlation energy in electronic structure calculations on infinite
systems of no particular symmetry. The vdW-DF functional [Phys.\ Rev.\ Lett.\
\textbf{92}, 246401 (2004)] is applied to the adsorption of benzene and
naphthalene on an infinite sheet of graphite, as well as the binding between
two graphite sheets.
Comparison with recent thermal desorption data [Phys.\ Rev.\ B\ \textbf{69},
535406 (2004)] shows great promise for the vdW-DF method. 
\end{abstract}

\maketitle

A recent study of the interaction of polycyclic aromatic hydrocarbon molecules 
(PAH's) with the basal plane of graphite \cite{zacharia} provides experimental 
benchmark data that constitute an ideal challenge for our recently proposed 
density functional (vdW-DF) \cite{gg} which both includes van der Waals (vdW) 
interactions and all the benefits of the earlier state-of-the-art versions 
of the density-functional theory (DFT). Aiming at a better experimental 
characterization of the weak interlayer interactions in graphite, careful 
analysis of thermal-desorption kinetics yield activation energies for benzene 
and PAH's at submonolayer coverages with explicit error bars \cite{zacharia}. 
Our calculated values for the adsorption energy of benzene and naphthalene on 
graphene and for the weak interlayer interaction energy of graphene agree 
with the values deduced from experiment. From this we conclude that the vdW-DF 
is, indeed, very promising, and that it can be applied to systems that are 
neither periodic nor finite. This distinguishes it from the various wave-%
function methods that are often applied to vdW complexes.

Our method differs also from a newly published study of adenine on graphite 
\cite{ortmann}, which treats the vdW part of the correlation energy by 
a frequently used \cite{scoles,kaxiras,weitao,hasegawa,zimmerli,grimme} semi-%
empirical method. This method introduces empirical damping functions applied 
to an asymptotic attractive $1/R^6$ interaction assumed to occur between each 
pair of nuclei. At shorter distances this interaction is damped by a physically 
motivated, but arbitrary and varying functional form, which introduces one 
empirical parameter for every pair of atomic types in the complex. On the other 
hand our method (vdW-DF) for the correlation energy is completely free from
empiricism, and although containing approximations, represents a first-principles
density functional, which since the appearance of Ref.~\onlinecite{gg} is 
applicable to arbitrary geometries, and which is seamless as two fragments 
merge into a single one. As discussed later, it has been applied to a number 
of physical systems with promising results. The present application is particularly 
pertinent, however, as alternative first-principles methods for including vdW 
interactions are lacking for extended systems.

Condensed matter is held together by several kinds of interatomic forces, including 
the ubiquitous vdW forces, which are particularly significant in sparse matter. For 
dense matter DFT has well proven its value, state-of-the-art versions of it giving 
values for ground-state properties of covalent molecules and hard materials close to 
experimental data. The key to success in DFT is the functional for exchange and 
correlation, and DFT calculations today typically apply some flavor of the generalized-%
gradient approximation (GGA) \cite{GGAmolecule,GGAsolid}. For sparse matter, including 
soft matter, vdW complexes, polymers, biomolecules, and other abundant systems, however, 
DFT in GGA performs badly. For instance, it gives unphysical results for the 
interlayer bond of graphite \cite{vdW3}, a canonical vdW case. The vdW interaction stems 
from truly nonlocal correlation and is contained in neither the local-density 
approximation (LDA) nor the semilocal GGA. We have subsequently proposed a density 
functional (vdW-DF) for general geometries \cite{gg} that includes both vdW interactions 
and all GGA achievements. It has been applied to the benzene dimer \cite{gg,aaron}
and to complexes of benzene with phenol, toluene, fluorobenzene, or benzonitrile 
\cite{timo}. All these previous applications have been to finite systems. We 
now apply it to single benzene and naphthalene molecules in interaction with 
an extended graphene sheet. This establishes the applicability of the vdW-DF 
functional to a more general class of systems. 

The derivation of the vdW-DF starts from the adiabatic connection formula and
benefits from insights into the polarization properties of the inhomogeneous 
electron gas. It involves several approximations, which satisfy known limits, 
sum rules, and invariances \cite{gg}. To assess the validity of vdW-DF, 
internal criteria must be supplemented with external ones, that is, we must 
evaluate its performance in actual applications by comparison with experimental 
findings. Systems where the vdW forces play a prominent role or dominate the 
interactions entirely are to be preferred. Being based on inhomogeneous-electron-%
gas polarization properties, the vdW-DF should work best on systems with similar 
polarization properties, that is, on delocalized systems with dense excitation 
spectra. Dimers of benzene and PAH's are such systems. Experimental data for these 
are scarce, and in practice the assessment has to be made by comparing to accurate 
wave-function calculations \cite{gg,aaron}, for which, however, the computational 
effort grows strongly with size, thus making benchmark results unavailable for 
really large systems. Further, to properly assess the vdW-DF method, extended 
systems should be addressed. The adsorption problem, with its extended substrate, 
is far beyond the possibilities of such wave-function methods. The agreement 
between values calculated with vdW-DF and measured values of key quantities is 
demonstrated below.

Recent thermal-desorption studies of the interaction of PAH's with the basal plane 
of graphite \cite{zacharia} yield, through the analysis of desorption kinetics, 
activation energies of 0.50, 0.85, 1.40, and 2.1 eV for benzene, naphthalene, 
coronene, and ovalene, respectively, at submonolayer coverages. Using a force-field 
model \cite{forcefield}, the authors of Ref.~\onlinecite{zacharia} predict an 
\textit{exfoliation energy} of $52\pm 5$ meV per/atom to peel a single graphene 
layer off a graphite surface. Another determination \cite{nanotubes} involving 
collapsed nanotubes, which obtained $35^{+15}_{-10}$ meV/atom 
for a related quantity (the energy to separate two graphene 
sheets), required an equally tenuous theoretical model to extract a prediction. 
Here we apply vdW-DF to calculate the binding energy of a benzene and a naphthalene
molecule to a single graphene sheet, as well as the binding energy of two graphene 
sheets to each other. We also present evidence that second-layer interactions are 
small, which imply that our results also apply approximately to molecular desorption 
or to graphene exfoliation from a basal graphite surface, as relevant for the 
experiment mentioned first above.

The general geometry (gg) vdW-DF \cite{gg} basically ``corrects'' the correlation 
part of the energy of a standard self-consistent (SC) DFT calculation, using the 
standard exchange from the GGA of the revPBE-flavor \cite{revPBE}, chosen for its 
being closest to the Hartree-Fock exchange in some key applications \cite{gg,aaron}. 
The vdW-DF energy functional is written 
\begin{equation}
  E_{\text{vdW-DF}} =  E_{\text{GGA}} - E_{\text{GGA,c}} 
+ E_{\text{LDA,c}} + E^{\text{nl}}_{\text{c}},
\label{eq:vdW-DF}
\end{equation}
where the LDA correlation energy $E_{\text{LDA,c}}$ is substituted for the GGA 
correlation energy, $E_{\text{GGA,c}}$, and where the nonlocal correlation energy 
$E^{\text{nl}}_{\text{c}}$ is added. The last three terms in (\ref{eq:vdW-DF}) are 
treated as a post-GGA perturbation, utilizing their low sensitivity to the choice of 
GGA SC electronic density. The nonlocal correlation energy is expressed as
\begin{equation}
E_{\text{c}}^{\text{nl}} = 
\frac{1}{2} \int\!d^3r\,d^3r' \, n(\vec r) \phi(\vec r,\vec r\,') n(\vec r\,'),
\label{eq:twopoint}
\end{equation}
where $\phi(\vec r,\vec r\,')$ is a function depending on $|\vec r -\vec r\,'|$, 
the charge density $n$ and its gradient at the positions $\vec r$ and $\vec r\,'$, 
respectively, given in detail by Eq.~(14) of Ref.~\onlinecite{gg}.

\begin{figure}
\includegraphics[width=.48\textwidth]{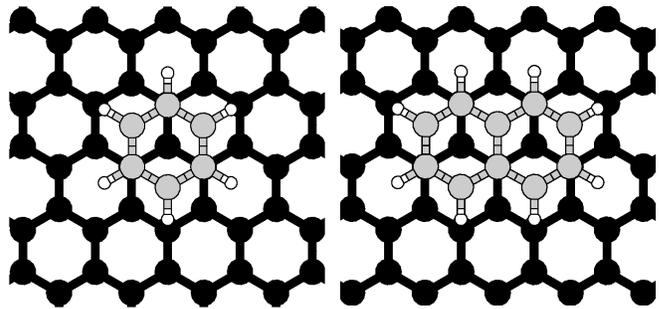}
\caption{The lateral configuration of the AB adsorption positions for the molecules 
benzene and naphthalene on a graphene sheet. Carbon atoms are shown in black, when 
belonging to the substrate and gray, when belonging to the adsorbate.
Hydrogen atoms are shown in white.}
\label{fig:adsorptiongeometry}
\end{figure}

SC calculations are performed with the 
pseudo\-potential-plane-wave-based 
DFT code \textsc{dacapo} \cite{dacapo}, using the revPBE functional
to get GGA electron densities and total energies. The hexagonal cell 
used for the standard part of the DFT calculations has the size (12.32 {\AA}, 
12.32 {\AA}, 26 {\AA}) for the benzene calculations and (14.783 {\AA}, 12.32 
{\AA}, 26 {\AA}) for the naphthalene calculations. We use a Monkhorst-Pack grid 
with (2, 2, 1) ${k}$-points and a plane-wave cutoff energy of 450 eV. The in-plane 
structure of the adsorbate is found through a monomer calculation, where the 
molecule is fully relaxed within a GGA-revPBE SC calculation. This results in an 
optimum structure with carbon-carbon and carbon-hydrogen bondlengths in agreement 
with experimental data \cite{SvetlaLicThesis}. In the same way, the structure of 
the graphene sheet is obtained from an isolated calculation, resulting in an 
optimum intralayer lattice constant of 2.46 {\AA}, in agreement with experiment \cite{Bornstein80}.
Afterwards the in-plane structures are kept fixed while 
we vary the distance from the adsorbate to the surface and map out an energy 
profile as a function of the separation. We place the molecule in an AB 
configuration (Fig.~\ref{fig:adsorptiongeometry}), which we expect to be 
energetically most favorable, as is the case for both benzene 
and naphthalene dimers, and for graphene layers. That is, we place 
the center of the benzene ring exactly above a carbon atom in the graphene sheet, 
and for naphthalene, the molecule center-of-mass is positioned above the center-of-mass of 
the two graphite atoms below the naphthalene rings, as seen in 
Fig.~\ref{fig:adsorptiongeometry}~\cite{NaphthaleneBondnote}. We know that for 
naphthalene dimers in the AB stacking, small shifts in the exact lateral position 
yield minor changes in the total energy, while the separation in the perpendicular 
direction is of much greater importance \cite{NaphDimerR1R2R3}, and we believe the 
same should hold also for small lateral shifts with respect to the graphene layer 
in Fig.~\ref{fig:adsorptiongeometry}.

For the adsorption position given above, full GGA calculations are performed, thus 
obtaining the first three terms in Eq.~(\ref{eq:vdW-DF}). The last term, given by Eq. 
(\ref{eq:twopoint}), requires some extra considerations, as (lateral) cell size becomes 
more important when evaluating the nonlocal vdW energy correction than for the standard 
calculations. For this purpose we enlarge the system size for the $E_{\text{c}}^{\text{nl}}$ 
evaluation as follows. The SC electron density for the adsorbate-substrate system obtained 
with GGA-revPBE is used within the unit cell, while outside this unit cell the graphene 
electronic density is simulated by that of pure graphene, enabling the extension of the 
substrate to, in principle, arbitrary size \cite{doublegrid}. Thus we can obtain the binding 
energy for benzene and naphthalene interacting with an increasingly larger circle of the 
graphene layer below the center of the adsorbed molecule. While 91\% of the interaction 
is obtained for the original unit cell, 99\% 
is obtained for a 10~{\AA} radius circle, and full convergence 
is reached for a 14~{\AA} circle. The same 
method is used for the calculation on the interactions of the two graphene layers. Thus 
calculation of $E_{\text{c}}^{\text{nl}}$ allows an increase of system size without much 
increased computational cost compared to standard DFT calculations.

\begin{figure}
\scalebox{0.40}{\includegraphics{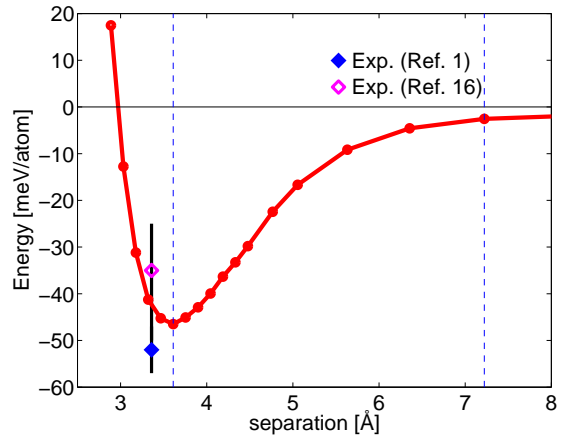}}
\caption{(Color online) Binding-energy curve for two graphene sheets in AB stacking. 
The calculated general geometry vdW-DF curve (solid line) indicates a binding energy of 
45.5 meV/atom at the equilibrium separation 3.6~{\AA}, which is close to the experimental 
energy estimates (diamonds with energy error bars; see text). The two vertical dashed 
lines mark the positions of the potential-energy minimum (3.6~{\AA}) and the approximate 
point where a next-neighbor graphite layer would have been placed (7.2~{\AA}). 
}
\label{fig:graphitesheetsAAAB}
\end{figure}

Binding-energy results for two interacting graphene sheets in the AB stacking, which is 
the structure occurring in bulk graphite, are shown in Fig.~\ref{fig:graphitesheetsAAAB}. 
The graphene--graphene binding-energy curve is somewhat deeper than that obtained with 
our earlier layered-functional version \cite{vdW3}. The general geometry version employs 
improvement made possible by an approximate expansion not made in the layered-functional 
version. Further discussion of the similarities and differences between the two methods 
is given separately \cite{svetla3}. The gg-vdW-DF curve deviates only slightly from the 
corresponding one in Ref.~\onlinecite{MaxThesis}, where an averaging procedure was used.

Our results for two graphene sheets in the AB stacking may be compared with the results 
of the experimental studies discussed above \cite{zacharia,nanotubes}. These are marked 
in Fig. \ref{fig:graphitesheetsAAAB} with diamonds at the layer separation in bulk graphite, 
which is $3.36$ {\AA}  \cite{Bornstein80}. We expect the separation of the AB graphene 
sheets to be similar, or only slightly larger. The energy error bars from the two 
experiments merge into one.

While the exfoliation energy from a graphite basal plane is not the same as the energy
to separate two graphene sheets, we may use the value of our curve at a second layer
distance to estimate the change in our prediction if an exfoliation calculation 
were done instead. From the value of our curve of $-2.5$ meV at 7.2 {\AA} in 
Fig.~\ref{fig:graphitesheetsAAAB}, we can estimate that our method will give a 
48 meV/atom exfoliation energy, and a 50.5 meV/atom graphite cleavage energy, a 5.5\% 
increase each ~\cite{girifalcoNote}. This conclusion requires the forces from subsequent 
layers to be negligible, which is certainly true for our functional, and probably also 
when the anomalous asymptotic effect of graphite's semimetallic nature \cite{dobson_asymptote} 
is accounted for. 
For graphite, thermal vibrations also give a relevant
contribution to the energy from the motion perpendicular to the sheets ~\cite{hasegawa}.
In any case the comparison of our curves with the points given by 
the two experimental groups is satisfying, despite the dependence of the latter on 
earlier theoretical methods that cannot be fully justified.

\begin{figure*}

\includegraphics[height=6cm]{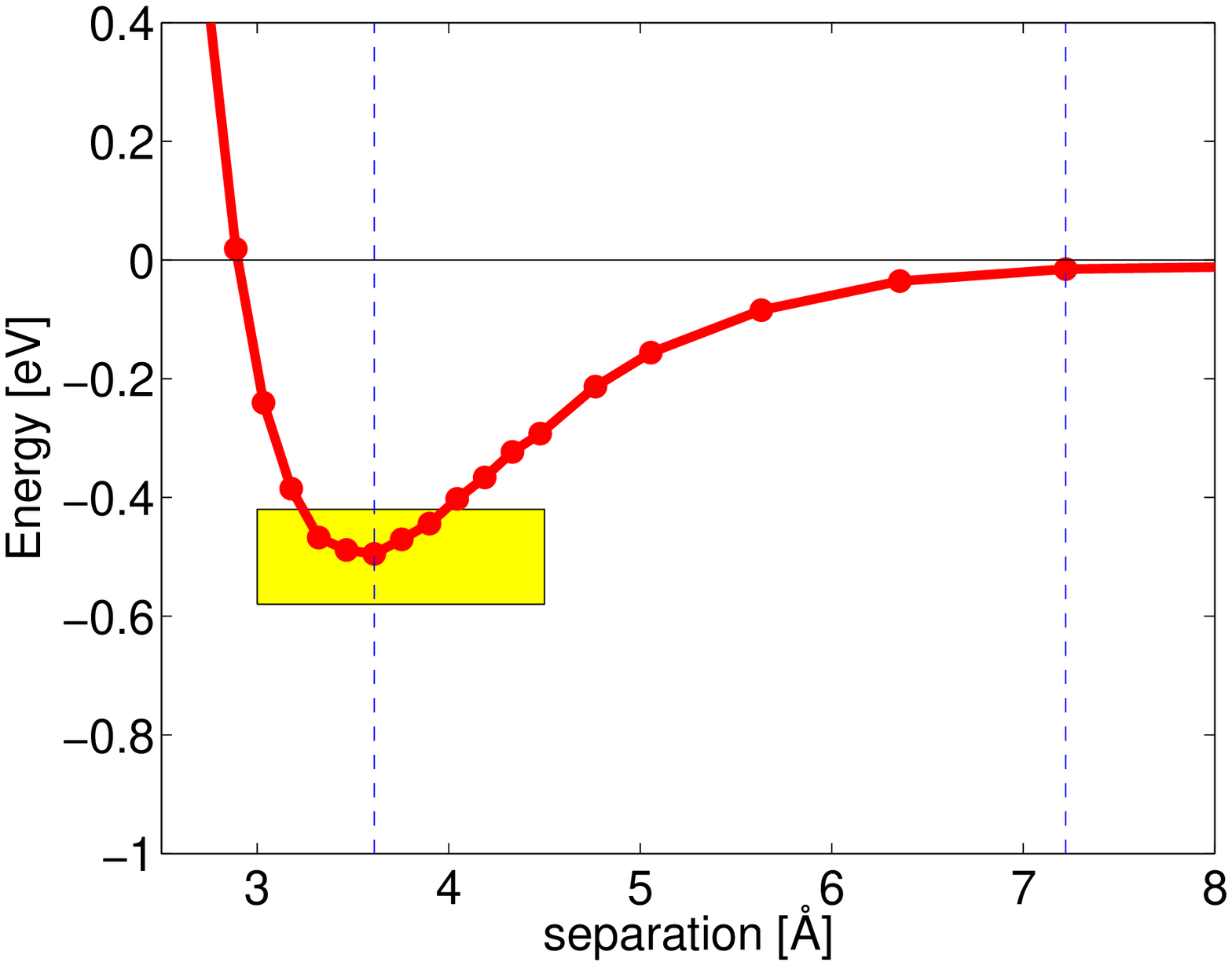}
\hspace{2em}
\includegraphics[height=6cm]{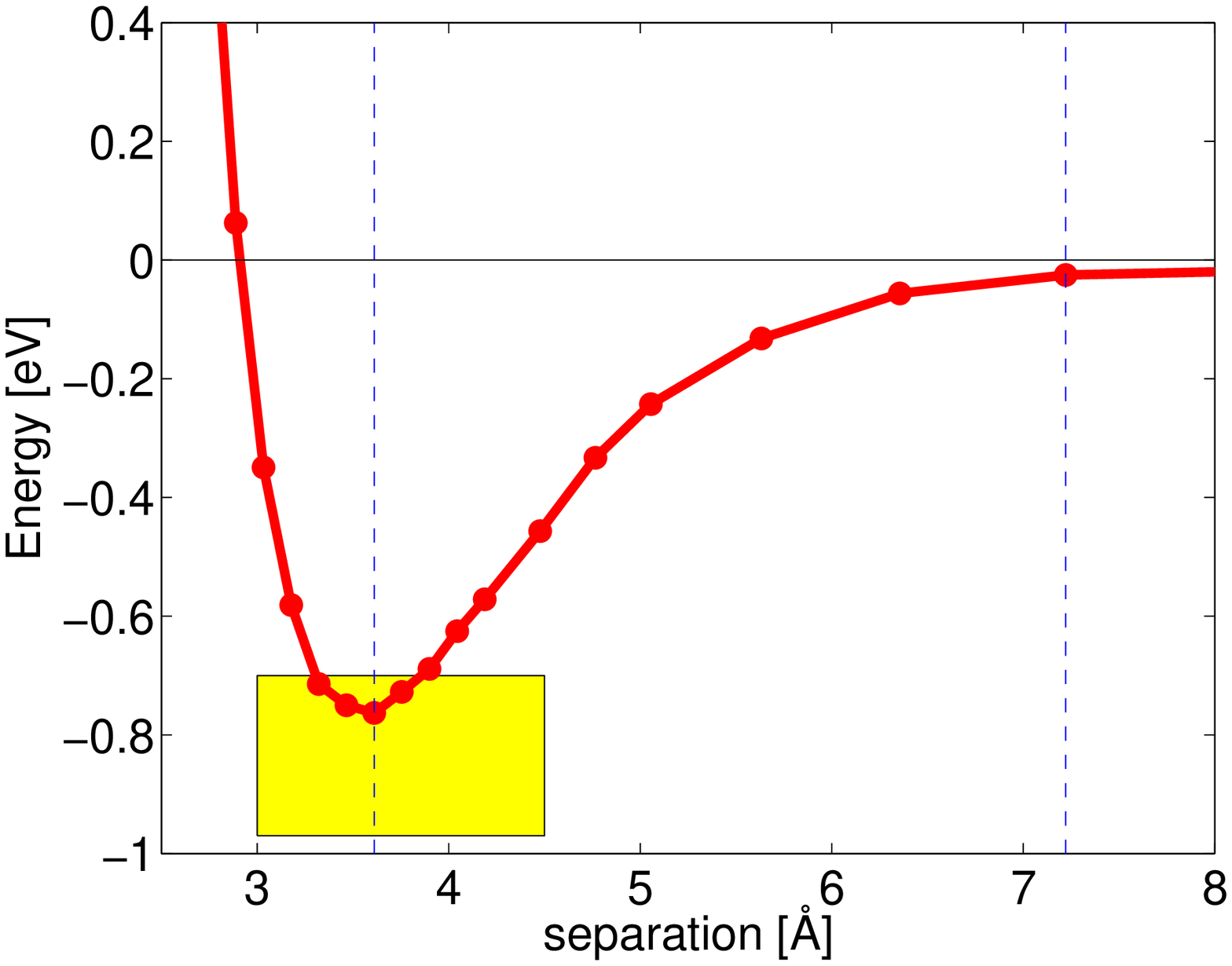}
\caption{(Color online) Benzene-graphene (left panel) and naphthalene-graphene (right panel) 
binding-energy curves. Like that for graphene-graphene, the general geometry vdW-DF 
calculations (solid lines) give results comparable to experiment (boxes); see text. The two 
vertical dashed lines mark the minima and the next-neighbor graphite layer, as in 
Figure~\protect~\ref{fig:graphitesheetsAAAB}.
}
\label{fig:B&NonG}
\end{figure*}

Binding-energy curves are also calculated for benzene and naphthalene adsorbed on graphene, 
with configurations as shown in  Fig.~\ref{fig:adsorptiongeometry}. We judge these 
calculations equivalent to those of adsorption on the basal plane of graphite, when a small 
correction for a second graphite layer (3.0\% increase for benzene and 3.3\% for naphthalene) 
is taken into account, as discussed above for graphene-graphene. Figure \ref{fig:B&NonG} 
shows the calculated gg-vdW-DF potential-energy curves of adsorbed benzene and naphthalene, 
respectively, with their minima well within the ranges of measured binding-energy values 
\cite{zacharia,pierce}. For benzene, we find a binding energy of 495 meV (to be compared to 
the values $500\pm80$ meV \cite{zacharia} and 480 meV \cite{pierce}) at the equilibrium 
separation 3.6~{\AA}. The same separation is found optimal for naphthalene, with the 
adsorption energy 763 meV, which (particularly, when including the second-layer 3.3\% 
contribution) can be compared to the experimental values $800\pm100$ meV and $900\pm70$ meV 
resulting from slightly different analysis of the same experiment \cite{zacharia}, summed up 
by the authors themselves by reporting the number $850$ meV. The experimental values with 
error bars on the energy are shown as shaded regions in the figures. The separations were not 
measured in the experiment. However, it is reasonable to expect the separation to be similar 
to that in graphite also in these two cases, which is what we find.

The analysis of the desorption experiments on benzene and naphthalene and the close 
agreement between experimental and theoretical results let us conclude that the vdW-DF is 
a very promising functional to account for the nonlocal vdW forces in these very typical 
cases. While the experimental evidence is not so direct in the graphene--graphene binding 
case, the values found are probably representative of what more direct experimental 
methods would find, and give additional support to the vdW-DF method.

Valuable exchanges with P. Hyldgaard and T. Hertel are gratefully acknowledged, 
as is support from the Swedish Foundation for Strategic Research via the ATOMICS consortium 
and the Swedish Research Council, as well as allocation of 
computer time at UNICC (Chalmers) and SNIC (Swedish National Infrastructure for 
Computing). Work by D.C.L.\ was supported in part by NSF Grant DMR-0456937.


\begin{thebibliography}{99}
\frenchspacing

\bibitem{zacharia}
R. Zacharia, H. Ulbricht, and T. Hertel,
Phys. Rev. B \textbf{69}, 155406 (2004).

\bibitem{gg}
M. Dion, H. Rydberg, E. Schr{\"o}der, D. C. Langreth, and B. I. Lundqvist,
Phys. Rev. Lett. \textbf{92}, 246401 (2004); \textbf{95}, 109902 (2005).

\bibitem{ortmann}
F. Ortmann, W.G. Schmidt, and F. Bechstedt,
Phys. Rev. Lett. \textbf{95}, 186101 (2005).

\bibitem{scoles}
X. Wu et al., 
J. Chem. Phys.
\textbf{115}, 8748 (2001).

\bibitem{kaxiras}
M. Elstner et al., 
J. Chem. Phys. \textbf{114}, 5149 (2001).

\bibitem{weitao}
Q. Wu and W. Yang, J. Chem. Phys. \textbf{116}, 515 (2002).

\bibitem{hasegawa}
M. Hasegawa and K. Nishidate, Phys. Rev. B \textbf{70}, 205431 (2004).

\bibitem{zimmerli}
U. Zimmerli, M. Parrinello, and P. Koumoutsakos, J. Chem. Phys. \textbf{120}, 2693 (2004).

\bibitem{grimme}
S. Grimme, J. Comp. Chem. \textbf{25}, 1463 (2004).

\bibitem{GGAmolecule}
V.~N. Staroverov, G.~E. Scuseria, J. Tao, and J. P. Perdew,
J. Chem. Phys. \textbf{119}, 12129 (2003).

\bibitem{GGAsolid}
V.~N. Staroverov, G.~E. Scuseria, J. Tao, and J. P. Perdew,
Phys. Rev. B \textbf{69}, 075102 (2004).

\bibitem{vdW3}
H. Rydberg, M. Dion, N. Jacobson, E. Schr{\"o}der, 
P. Hyldgaard, S. I. Simak, D. C. Langreth, and B. I. Lundqvist,
Phys. Rev. Lett.\ \textbf{91}, 126402 (2003).

\bibitem{aaron}
A. Puzder, M. Dion, and  D. C. Langreth,
cond-mat/0509421 (submitted to J. Chem. Phys.).

\bibitem{timo}
T. Thonhauser, A. Puzder, and D. C. Langreth,
cond-mat/0509426 (submitted to J. Chem. Phys.).

\bibitem{forcefield}
N. L. Allinger, Y. H. Yuh, and J.-H. Lii,
J. Am. Chem. Soc. \textbf{111}, 8551 (1989).

\bibitem{nanotubes}
L.X. Benedict et al., 
Chem. Phys. Lett. \textbf{286}, 490 (1998). 

\bibitem{revPBE}
Y. Zhang and W. Yang,  
Phys. Rev. Lett. \textbf{80}, 890 (1998). 

\bibitem{dacapo}
Computer code \textsc{dacapo}, 
\texttt{http://www.fysik.dtu.dk /CAMPOS/}

\bibitem{SvetlaLicThesis}
S. Chakarova, Licenciate Thesis, 
Chalmers University of Technology, G{\"o}teborg, Sweden (2004). 

\bibitem{Bornstein80}
{\it Landolt-B{\"o}rnstein search} (Springer, Berlin, 2003), 
http://link.springer.de.

\bibitem{NaphthaleneBondnote}
The C--C bondlengths in naphthalene differ from those in graphite by
up to 3\%.

\bibitem{NaphDimerR1R2R3}
S. Tsuzuki, K. Honda, T. Uchimaru, and M. Mikami,
J. Chem. Phys. \textbf{120}, 647 (2004). 

\bibitem{doublegrid} 
For $E^{\text{nl}}_{\text{c}}$ we sampled the electron density
only on every second standard-DFT Fast-Fourier-Transform (FFT) gridpoint in each direction.
(Our standard-DFT calculations have approx. 0.14 {\AA}/FFT-gridpoint). 
A finer sampling scheme using every
FFT gridpoint was tested and found unnecessary. 

\bibitem{svetla3}
S. D. Chakarova K\"ack, J. Kleis, and E. Schr\"oder, 
\textit{Dimers of polycyclic aromatic hydrocarbons in density functional theory},
to be published.

\bibitem{MaxThesis}
M. Dion, Ph.D. Thesis, 
Rutgers University, New Brunswick, New Jersey, (2004). 

\bibitem{girifalcoNote}
Earlier theory [L. A. Girifalco and R. A. Lad, J. Chem. Phys. \textbf{25}, 693 (1956)] 
implied a larger 18\% 
difference, later used in \cite{zacharia} 
to predict a cleavage energy of 61 meV/atom. 

\bibitem{dobson_asymptote}
J. F. Dobson, A. White, and A. Rubio, cond-mat/0507156.

\bibitem{pierce}
C. Pierce, Journ. Phys. Chem. \textbf{73}, 813 (1969).

\end{thebibliography}
\end{document}